\documentstyle[11pt,multicol,epsfig]{article}

\begin{document}
\input epsf

\title{ Dynamical Neural Network: Information and Topology }

\author{ David Dominguez$$
\thanks{DD thanks a Ramon y Cajal grant from MCyT.
E-mail: david.dominguez@uam.es
} ,
Kostadin Koroutchev$$ \\
Eduardo Serrano$$ and 
Francisco B. Rodr\'{\i}guez \\
EPS, Universidad Autonoma de Madrid, 
Cantoblanco, Madrid, 28049, Spain 
}

\date{\today}
\maketitle


\begin{abstract}
A neural network works as an associative memory device 
if it has large storage capacity and the quality of the
retrieval is good enough.
The learning and attractor abilities of the network both
can be measured by the mutual information (MI),
between patterns and retrieval states.
This paper deals with a search for an optimal topology,
of a Hebb network, in the sense of the maximal MI.
We use small-world topology.
The connectivity $\gamma$
ranges from an extremely diluted 
to the fully connected network;
the randomness $\omega$ ranges from purely local
to completely random neighbors.
It is found that,
while stability implies an optimal $MI(\gamma,\omega)$ at
$\gamma_{opt}(\omega)\to 0$,
for the dynamics,
the optimal topology holds at certain
$\gamma_{opt}>0$ whenever $0\leq\omega<0.3$.
\end{abstract}


\section{Introduction}

The collective properties of attractor neural networks 
({ANN}),
such as the ability to perform as an associative memory,
has been a subject of intensive research in the last couple of
decades\cite{HK91},
dealing mainly with fully-connected topologies.
More recently, 
the interest on ANN has been renewed by the study of
more realistic architectures, 
such as small-world \cite{SW98},\cite{MA04}
or scale-free \cite{AB02},\cite{To04} models.
The storage capacity $\alpha_c$ and the overlap $m$ 
with the memorized patterns 
are the most used measures of the retrieval ability
for the Hopfield-Hebb networks\cite{AG87},\cite{Ok96}.
Comparatively less attention has been paid to the
study of the mutual information (MI) between stored patterns
and the neural states\cite{PA89}\cite{DB98},
although neural networks are information processing machines.

A reason for this relatively low interest is twofold:
on the one hand, 
it is easier to deal with the global parameter
$m[\vec{\sigma},\vec{\xi}]$,
than with $MI[p(\vec{\sigma}|\vec{\xi})]$,
a function of the conditional probability of neuron states $\vec{\sigma}$ 
given the patterns $\vec{\xi}$.
This can be solved for the so called {\em mean-field networks} 
which satisfy the law of large numbers, 
hence MI is a function only of the macroscopic parameters $m$, 
and the load rate $\alpha=P/K$ 
(where $P$ is the number of uncorrelated patterns,
and $K$ is the neuron connectivity).
On the other hand, 
the load $\alpha$ is enough to measure the information
if the overlap is close to $m\sim 1$,
since in this case the information carried by any single 
binary neuron is almost 1 bit.
It is true for a fully-connected (FC) network,
for which the critical $\alpha^{FC}_c\sim 0.138$ \cite{AG87}, 
with $m^{FC}_c\sim 0.97$ 
(with a sharp transition to $m\to 0$ for larger $\alpha\geq\alpha_c$):
in this case, the information rate is about $i^{FC}_c\sim 0.131$,
as can be seen in the left panel of Fig.\ref{im,ac}.
There we show the overlap (upper)
and information for several architectures.
However, in the case of diluted networks
the transition is smooth.
In particular, 
the random extremely diluted (RED) network
has load capacity
$\alpha^{RED}_c\sim 0.64$\cite{CG88} 
but the overlap falls continuously to $m^{RED}_c\sim 0$,
which yields null information at the transition,
$i^{RED}_c\sim 0.0$,
as seen in right panel of Fig.\ref{im,ac} 
(dashed line).
Such indetermination shows that one must search for the
value of $\alpha_{max}$ corresponding to the maximal information
$MI_{max}\equiv MI(\alpha_{max})$,
instead of $\alpha_c$.

\begin{figure}[t]
\begin{center}
\epsfxsize 11.cm \epsfysize 8.cm
\epsfbox{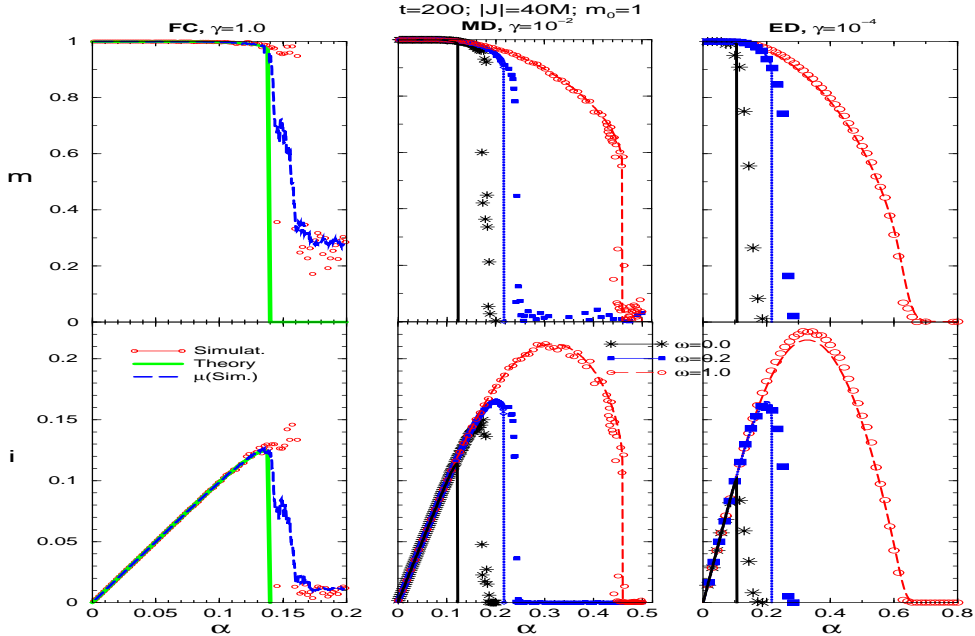}
\caption{ {\it \small{
The overlap $m$ and the information $i$ vs $\alpha$ for 
different architectures: 
fully-connected, $\gamma^{FC}=1.0$ (left),
moderately-diluted, $\gamma^{MD}=10^{-2}$ (center)
and extremely-diluted, $\gamma^{ED}=10^{-4}$ (right).
Symbols represents simulation with initial overlap $m^0=1$
and $|J|=40M$, with
local (stars, $\omega=0.0$),
small-world (filled squares, $\omega=0.2$),
and random (circles, $\omega=1.0$) connections.
Lines are for theoretical results:
solid, $\omega=0.0$,
dotted, $\omega=0.2$,
and dashed, $\omega=1.0$.
In left, dashed line means averaging the simulation.}
} }
\label{im,ac}
\end{center}
\end{figure}

We address the problem of searching for the optimal topology,
in the sense of maximizing the mutual information.
Using the graph framework \cite{AB02},
one can capture the main properties of a wide range of
neural systems,
with only 2 parameters:
$\gamma\equiv K/N$, which is the average rate of links per neurons, 
where $N$ is the network size,
and $\omega$, which controls the rate of random links 
(among all neighbors).
When $\gamma$ is large,
the clustering coefficient is large ($c\sim 1$) and the 
mean-length-path between neurons is small ($l\sim \ln N$), 
whatever $\omega$ is.
When $\gamma$ is small,
then if $\omega$ is too small, 
$c\sim 1$ and $l\sim N/K$,
but if it is about $\omega\sim 0.1$,
the network behaves again as if $\gamma\sim 1$,
with $c\sim 1$ and $l\sim \ln(N)$.
This region, called small-world (SW), 
is rather usefull when one is interested to built
networks where the information transmition is fast and efficient,
with high capacity in presence of significant noise,
but do not wants to spent too much wiring \cite{AC05}. 
Small-world networks may model many biological systems \cite{Sp04}.
For instance, 
in a brain local connections dominate in intracortex,
while there are a few intercortical connections \cite{RT04}.

In Fig.\ref{im,ac} we show the overlap (upper)
and information for several architectures.
In the left panel,
it is seen that the maximum information rate,
$i\equiv MI/(K.N)$, of FC network is about 
$i_{max}^{FC}=0.135$,
while in the right panel,
we show extremely-diluted networks (ED).
The RED network ($\omega=1.0$) has $i_{max}^{RED}\sim 0.223$.
The right panel of Fig.\ref{im,ac} plot also the overlap and the 
information for the local extremely diluted network 
(LED, $\omega=0.0$),
with $i_{max}^{LEC}=0.0855$,
and a small-world extremely diluted network 
(SED, $\omega=0.2$),
with $i_{max}^{SED}=0.165$. 
We see that the ED transitions are smooth.
The central panel of Fig.\ref{im,ac} 
plot moderately diluted (MD) networks,
which are commented later.
Theoretical results fit well with the simulations,
except for small $\omega$,
where theory underestimate it.
Previous works about small-world
attractor neural networks \cite{MM03}  
studied only the overlap $m(\alpha)$,
so no result about information were known.

Our main goal in this work is to solve the following question:
how does the maximal information,
$i_{max}(\gamma,\omega)
\equiv i(\alpha_{max};\gamma,\omega)$ 
behaves with respect to the network topology?
To our knowledge, up to now,
there were no answer to this question.
We will show that, 
near to the stationary retrieval states,
for every value of the randomness $\omega>0$,
the extremely-diluted network,
performs the best, $\gamma_{opt}\to 0$.
However, regarding the attractor basins,
starting far from the patterns,
the optimal topology holds for moderate $\gamma_{opt}$.
For instance, if transients are taken in account,
values of $\omega\sim 0.1$ lead to an optimal 
$i_{opt}(\gamma)\equiv i_{max}(\gamma_{opt},\omega)$ 
with $\gamma_{opt}\sim 10^{-2}$.

The structure of the paper is the following:
in the next section we review the information measures
used in the calculations; 
in Sec.3, we define the topology and neuro-dynamics model.
The results are shown in Sec.4, 
where we study retrieval by theory and
simulation (with random patterns
and with images);
conclusions are drawn in last section.

\section{The Information Measures}

\subsection{The Neural Channel}

The network state at a given time $t$ is defined by 
a set of binary neurons,
$\vec{\sigma}^{t}=\{\sigma_i^t\in\{\pm 1\},i=1,...,N\}$.
Accordingly, each pattern
$\vec{\xi}^{\mu}=\{\xi_i^{\mu}\in\{\pm 1\},i=1,...,N\}$,
is a set of site-independent random variables,
binary and uniformly distributed:
$p(\xi^{\mu}_{i}=\pm 1)= 1/2$.
The network learns a set of independent patterns
$\{\vec{\xi}^{\mu},\,\,\mu=1,...,P\}$.

The task of the neural channel is to retrieve a pattern 
(say, $\vec{\xi}^{}$) starting from a neuron state which is 
inside its attractor basin, $B(\vec{\xi}^{})$, {\it i.e.}:
$\vec{\sigma}^{0}\in B(\vec{\xi}^{})\rightarrow
\vec{\sigma}^{\infty}\approx \vec{\xi}^{}$.
This is achieved through a network dynamics,
which couples neighbor neurons $\sigma_i,\sigma_{j}$ by the
{\em synaptic matrix} 
${\bf J}\equiv\{J_{i,j}\}$ 
with cardinality 
$|{\bf J}|=N\times K$.

\subsection{The Overlap}

For the usual binary non-biased neurons model,
the relevant order parameter is the $overlap$ between the
neural states and a given pattern:

\begin{equation}
m^{\mu t}_{N}\equiv
{1\over N}\sum_{i}\xi^{\mu}_{i}\sigma_{i}^t,
\label{3.mm}
\end{equation}
at the time step $t$.
Note that both positive $\bf{\xi}$
and negative $-\bf{\xi}$ patterns,
carry the same information,
so the absolute value of the overlap measures
the retrieval quality:
$|m|\sim 1$ means a good retrieval.
Alternatively,
one can measure the error in retrieving using the 
Hamming distance:
$D^{\mu t}_{N} \equiv
{1\over N}\sum_{i}|\xi^{\mu}_{i}-\sigma_{i}^t|^{2}
=2(1-m^{\mu t}_{N})$.

Together with the overlap,
one needs a measure of the load,
which is the rate of pattern bits per synapses used to store them.
Since the synapses and patterns are independent,
the load is given by 
$\alpha=|\{\vec{\xi}^{\mu}\}|/|{\bf J}|=(PN)/(NK)=P/K$.

We require our network to have long-range interactions.
Therefore, we regard a mean-field network (MFN),
the distribution of the states is site-independent,
so every spatial correlation such as 
$\langle\sigma_{i}\sigma_{j}\rangle-
\langle\sigma_{i}\rangle\langle\sigma_{j}\rangle$
can be neglected,
which is reasonable in the asymptotic limit
$K,N\to\infty$.
Hence the condition of the law of large numbers,
are fulfilled.
At a given time step of the dynamical process,
the network state can be described by one particular overlap,
let say $m_N^t\equiv m^{\mu t}_{N}$.
The order parameters can thus be written, 
when $N\to\infty$, as 
$m^t= \langle
\sigma^t{\xi}
\rangle_{\sigma,\xi}$.   
The brackets represent average over the joint distribution 
$p(\sigma|\xi)$, 
for a single neuron
(we can drop the index $i$).
This macroscopic variable describes the information processing
of the network, 
at a given time step $t$ of the dynamics.
Along with this signal parameter,
the residual $P-1$ microscopic overlaps yield the cross-talk noisy,
its statistics complete the network macro-dynamics.

\subsection{Mutual Information}

For a long-range system,
it is enough to observe the distribution of a single
neuron in order to know the global distribution \cite{DB98}.
This is given by the conditional probability of
having the neuron in a state $\sigma$,
at each (unspecified) time step $t$,
given that in the same site the pattern being retrieved
is $\xi$.
For the binary network we are considering, 
$p(\sigma|\xi)=(1+m\sigma\xi)\delta(\sigma^2-1)$, \cite{BD00}
where the overlap is
$m= \langle\langle\sigma 
\rangle_{\sigma|\xi} \xi \rangle_{\xi}$.

The joint distribution of $p(\sigma,\xi)$
is interpreted as an ensemble distribution
for the neuron states $\{\sigma_{i}\}$ 
and inputs $\{\xi_{i}\}$.
In the conditional probability,
$p(\sigma|\xi)$,
all type of noise in the retrieval process of the input pattern
through the network
(both from environment and over the
dynamical process itself) is enclosed.

With the above expressions and
$p(\sigma)\equiv\sum_{\xi}p(\xi)p(\sigma|\xi)=
\delta(\sigma^2-1)$,
we can calculate the MI \cite{DB98},
a quantity used to measure the prediction that an observer 
at the output ($\bf{\sigma}$) can do about the input 
($\bf{\xi}^{\mu}$) (we drop the time index $t$).
It reads 
$MI[\sigma;\xi]=S[\sigma]-S[\sigma|\xi]$,
where $S[\sigma]$ is the entropy and $S[\sigma|\xi]$ is
the conditional entropy.
We use binary logarithms to
measure the information in bits.
The entropies are \cite{BD00}:
\begin{eqnarray}
&&
S[\sigma|\xi]
= 
-{1+m\over 2}\log_2{1+m\over 2} - 
{1-m\over 2}\log_2{1-m\over 2} , \,\,
\nonumber\\
&& S[\sigma]=1[bit].
\label{3.He}
\end{eqnarray}

We define the information rate as 
\begin{equation}
i(\alpha,m)=MI[\vec{\sigma}|\{\vec{\xi}\mu\}]/|{\bf J}|\equiv
\alpha MI[\sigma;\xi],
\label{3.ia}
\end{equation}
since for independent neurons and patterns,
$ MI[\vec{\sigma}|\{\vec{\xi}\mu\}] \equiv
\sum_{i\mu} MI[\sigma_i|\xi^{\mu}_{i}] $.
When the network approaches its saturation limit $\alpha_c$,
the states can not remain close to the patterns,
then $m_c$ is usually small.
So, while the number of patterns increase,
the information per pattern decreases.
Therefore,
information $i(\alpha,m)$ is a non-monotonic function of
the overlap and load rate, 
see Fig.\ref{im,ac}, 
which reaches its maximum value $i_{max}=i(\alpha_{max})$
at some value of the load $\alpha_{max}$.

\section{The Model}

\subsection{The Network Topology}

The synaptic couplings are 
$J_{ij}\equiv C_{ij}W_{ij}$,
where the connectivity matrix has a local and a random parts,
$\{C_{ij}= C^n_{ij}+C^r_{ij} \}$,
and {\bf W} are synaptic weights.
The local part connects the $K_n$ nearest neighbors,
$C^n_{ij}=\sum_{k\in V} \delta(i-j-k)$,
with $V=\{1,...,K_n\}$ in the asymmetric case,
on a closed ring.
The random part consists of independent random variables 
$ \{C^r_{ij}\}$, distributed 
with probability 
$p(C^r_{ij}=1)=c_r$, and $C^r_{ij}=0$ otherwise,
with $c_r=K_r/N$, where $K_r$ is the mean number of
random connections of a single neuron.
Hence, the neuron connectivity is
$K=K_n+K_r$.
The network topology is then characterized by two parameters:
the {\it connectivity} ratio, defined as $\gamma=K/N$,
and the {\it randomness} ratio, $\omega=K_r/K$.
The $\omega$ plays the role of rewiring probability
in the {\em small-world} model (SW) \cite{SW98}.
Our model was proposed by Newman and Watts \cite{NW99},
which has the advantage of avoiding disconneting the graph.

Note that the topology  ${\bf C}$ can be defined by an 
adjacency list connecting neighbors, 
$i_k, k=1,...,K$, with $C_{ij}=1:j=i_k$.
So the storage cost of this network is $|{\bf J}|=N\cdot K$.
Hence, the information is $i=\alpha MI$, Eq.(\ref{3.ia}),
where the load rate is scaled as $\alpha=P/K$.
The learning algorithm updates {\bf W},
according to the Hebb rule
\begin{equation}
W_{ij}^{\mu}= W_{ij}^{\mu-1}+{1\over K}
\xi_{i}^{\mu} \xi_{j}^{\mu}.
\label{2.Ki}
\end{equation}
The network starts at $W^{0}_{ij}=0$,
and after $\mu=P=\alpha K$ learning steps,
it reaches a value 
$W_{ij}={1\over K}\sum_{\mu}^{p}
\xi_{i}^{\mu} \xi_{j}^{\mu}$.
The learning stage is a slow dynamics,
being stationary-like in the time scale of the 
much faster retrieval stage,
we define in the following.

\subsection{The Neural Dynamics}

The neural states, $\sigma_i^t\in\{\pm 1\}$,
are updated according to the stochastic parallel dynamics:
\begin{eqnarray}
\sigma^{t+1}_i={\rm sign}(h_i^t+Tx), \,\,
h_{i}^t \equiv \sum_{j} J_{ij} \sigma_j^t ,\,\,i=1...N
\label{2.st}
\end{eqnarray}
where $x$ is a normalized random variable
and $T$ is the temperature-like environmental noise.
In the case of symmetric synaptic couplings, $J_{ij}=J_{ji}$,
an energy function 
$H_{s} = -\sum_{(i,j)} J_{ij} \sigma_i \sigma_j$
can be defined,
whose minima are the stable states of the dynamics 
Eq.(\ref{2.st}).

In the present paper, 
we work out the asymmetric network by simulation
(no constraints $J_{ij}=J_{ji}$).
The theory was carried out for symmetric networks.
As it is seen in Fig.\ref{im,ac},
theory and simulation shows similar results,
except for local networks 
(theory underestimate $\alpha_{max}$,
where the symmetry may play some role. 
We restrict our analysis also for the deterministic dynamics
($T=0$).
The stochastic macro-dynamics comes from the extensive number
of learned patterns, $P=\alpha K$.

\section{Results}

\begin{figure}[t]
\begin{center}
\begin{minipage}{6.0cm}
\epsfxsize 6.cm \epsfysize=7.cm
\epsfbox{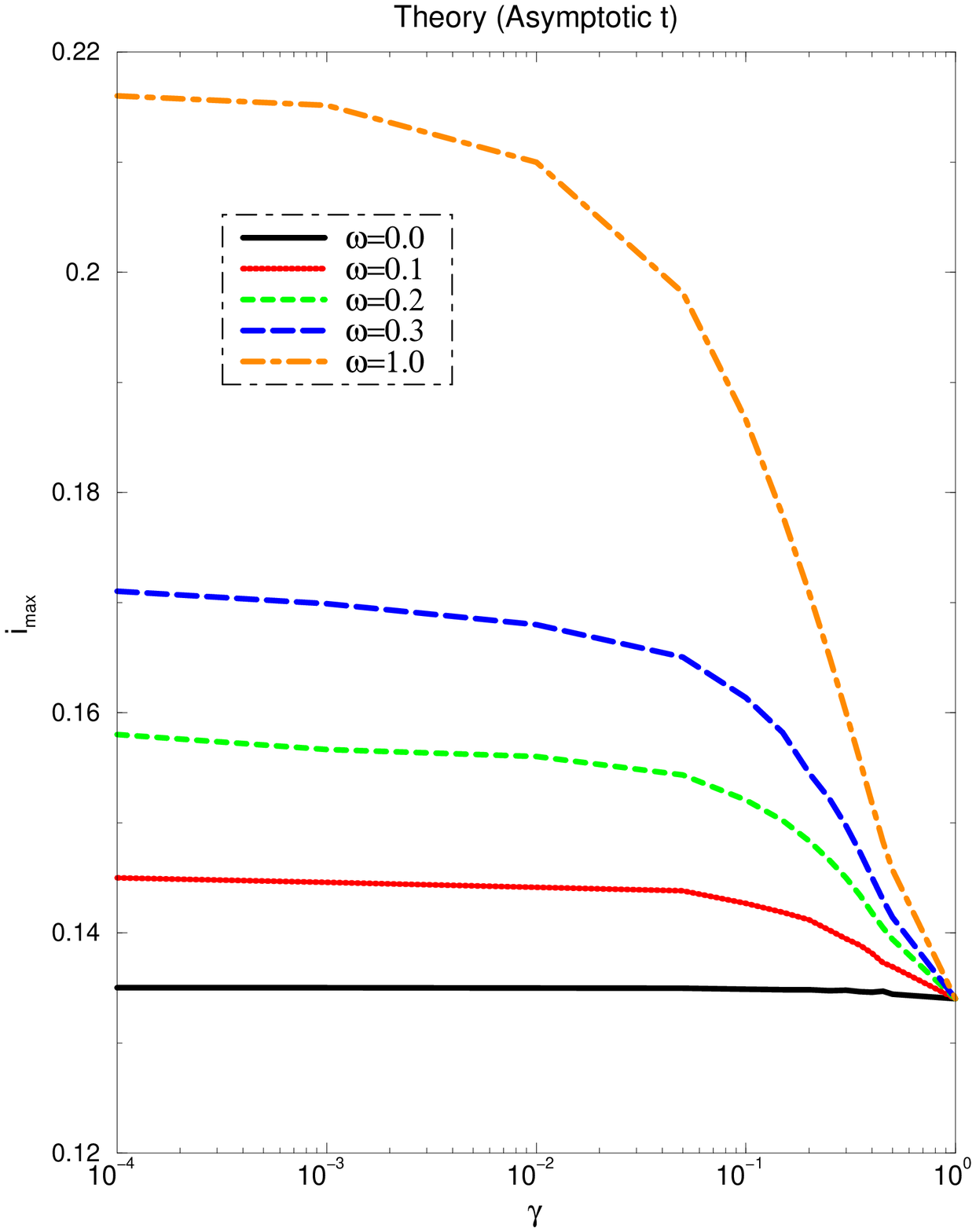}
\caption{\small
Maximal information $i_{max}=i(\alpha_{max})$ vs $\gamma$.
Theoretical results for the stationary states,
with several values of randomness $\omega$.}
\label{i,gwT}
\end{minipage}
\hfill
\begin{minipage}{6.cm}
\epsfxsize 6.cm \epsfysize=7.cm
\epsfbox{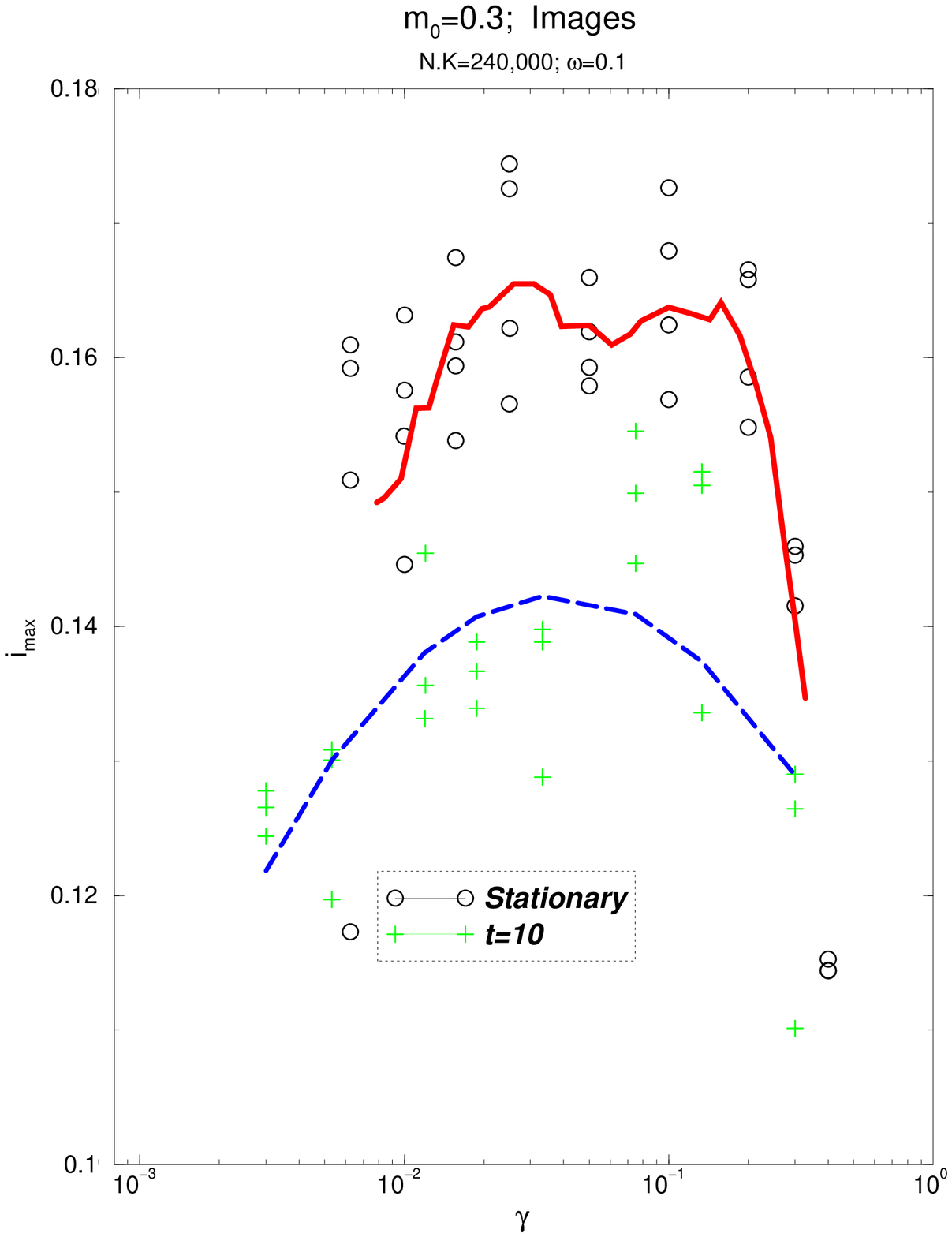}
\vspace{-0.6cm}
\caption{\small
$i_{max}$ vs $\gamma$.
Simulation with 
$\omega=0.1$ and $m^0=0.3$.
Dynamics stop at $t=10$ (Plus dots, Solid line)
or at $t=100$ (Circles, Dashed line).
}
\label{i,gw1}
\end{minipage}
\end{center}
\end{figure}

We studied the information for the stationary and dynamical states 
of the network were studied
as a function of the topological parameters,
$\omega$ and $\gamma$.
A sample of the results for simulation and theory is shown
in Fig.\ref{im,ac},
where the stationary states of the overlap and information are plotted
for the FC, MD and ED arquitetures.
It can be seen that information increases with dilution and with
randomness of the network.
A reason for this behavior is that dilution decreases the correlation
due to the interference between patterns.
However, dilution also increases the mean-path-length of the network,
thus, if the connections are local,
the information flows slowly over the network.
Hence, the neuron states can be eventually trapped in noisy patterns.
So, $i_{max}$ is small for $\omega\sim 0$ even if $\gamma=10^{-4}$.

\subsection{Theory: Stationary States}

Following to the Gardner calculations\cite{CG88},
at temperature T=0 the MFN approximation gives
the fixed point equations:
\begin{eqnarray}\label{ww}
  && m = {\rm erf}(m/\sqrt{r\alpha}),\,\,  \\
  && \chi = 2\varphi(m/\sqrt{r\alpha})/\sqrt{r\alpha};\\
  && r = \sum_{k=0}^\infty a_k (k+1) \chi^{k},
  \,\;a_k=\gamma Tr[({\bf C}/K)^{k+2}]
\end{eqnarray}
with ${\rm erf}(x)\equiv 2\int_0^{x} \varphi(z) dz $,
$\varphi(z)\equiv e^{-z^2/2}/\sqrt{2\pi} $.
The parameter $a_k$ 
is the probability of existence
of cycle of length $k+2$ in the connectivity graph.
The $a_k$ can be calculated either by using Monte Carlo \cite{DK04},
or by an analytical approach,
which gives $a_k\sim \sum_m \int d\theta [p(\theta)]^k e^{im\theta}$,
where $p(\theta)$ is the Fourier transform of the probability of
links, $p(C_{ij})$.
For an RED and FC networks one recover the known results
for $r^{RED}=1$ and $r^{FC}=1/(1-\chi)^2$ respectively \cite{HK91}.

The theoretical dependence of the information on the load,
for FC, MD and ED networks,
with local, small-world and random connections, 
are plotted in the fat lines in Fig.\ref{im,ac}.
A comparison between theory and simulation is also given
in Fig.\ref{im,ac}.
It can be seen that both results agree for most $\omega>0$,
but theory fails for $\omega=0$.
One reason is that theory uses symmetric constraint,
while simulation was carried out with asymmetric synapsis.
Figure \ref{i,gwT} shows their maxima $i(\alpha_{max})$ vs. 
the parameters $(\omega,\gamma)$. 
It is seen that the optimal is at $\omega\to 1, \gamma\to 0$.
This implies that the best topology for information (stationary states)
is the extreme diluted network,
with purely random connectivity.

\subsection{Simulation: Attractors and Transients}

We have studied the behavior of the network
varying the range of connectivity $\gamma$ and randomness $\omega$.
We used Eq.(\ref{2.st}). 
Both local and random connections are asymmetric.
The simulation was carried out with 
$N\times K=36\cdot 10^{6}$ synapses,
storing an adjacency list as data structure,
instead of $J_{ij}$.
For instance, with $\gamma\equiv K/N=0.01$,
we used $K=600,N=6\cdot 10^{4}$.
In \cite{MM03} the authors use $K=50,N=5\cdot 10^{3}$,
which is far from asymptotic limit.

We studied the network by searching
for the stability properties and transients of the neuron dynamics.
To look for stability, 
we started the network at some pattern (with initial overlap $m^0=1.0$),
and wait until it stays or leave it after a flag time step $t=t_f$
(unless it converges to a fixed point $m^*$ before $t=t_f$).
When we check transients, 
we start with $m^0=0.1$,
and stop the dynamics at the time $t_f$. 
Usually, 
$t_f=20$ parallel (all neurons) updates
is a large enough delay for retrieval.
Indeed in most case far before the saturation,
after $t_f=4$ the network end up in a pattern,
however, near $\alpha_{max}$,
even after $t_f=100$ the network has not yet relaxed.

In first place,
we checked for the stability properties of the network:
the neuron states start precisely at a given pattern
$\vec{\xi}^{\mu}$
(which changes at each learned step $\mu$).
The initial overlap is $m_0^{\mu}=1.0$,
so, after $t_m\leq 20$ time steps in retrieving,
the information $i(\alpha,m;\gamma,\omega)$
for final overlap is calculated.
We plot it as a function of $\alpha$,
and its maximum $i_{max}\equiv i(\alpha_{max};\gamma,\omega)$
is evaluated.
We averaged over a window in the axis of $P$,
usually $\delta P=25$.
This is repeated  for various values of the
connectivity ratio $\gamma$
and randomness $\omega$ parameters.
The results are in the upper panels of Fig.\ref{i,agm}.

Second,
we checked for the retrieval properties:
the neuron states start far from a learned pattern,
but inside its basin of attraction,
$\vec{\sigma}^{0}\in B(\vec{\xi}^{\mu})$.
The initial configuration is chosen with distribution:
$p(\sigma^0=\pm\xi^{\mu}|\xi^{\mu})=(1\pm m^0)/2$,
for all neurons
(so we avoid a bias between local/random neighbors).
The initial overlap is now $m^{0}=0.1$,
and after $t_f\leq 20$ steps,
the information $i(\alpha,m;\gamma,\omega)$
is calculated.
The results are in the lower panels of Fig.\ref{i,agm}.
The first observation now is that the maximal
information $i_{max}(\gamma;\omega)$ increases with
dilution (smaller $\gamma$) if the network is more random,
$\omega\simeq 1$,
while it decreases with dilution if the network is more local,
$\omega\simeq 0$.

The comparison between upper ($m^{0}=1.0$) and lower
parts of Fig.\ref{i,agm},
shows that the non-monotonic behavior of the information with
dilution and randomness,
is stronger for the retrieval ($m^{0}=0.1$) than for the
stability properties ($m^{0}=1.0$).
One can understand this in terms of the basins of attraction.
Random topologies have very deep attractors,
specially if the network is diluted enough,
while regular topologies almost lose their retrieval abilities
with dilution.
However, since the basins becomes rougher with dilution,
then network takes longer to reach the attractor.
Hence, the competition between depth-roughness is won by
the more robust MD networks.

\begin{figure}[t]
\begin{center}
\epsfxsize 11.cm
\epsfysize 8.cm
\epsfbox{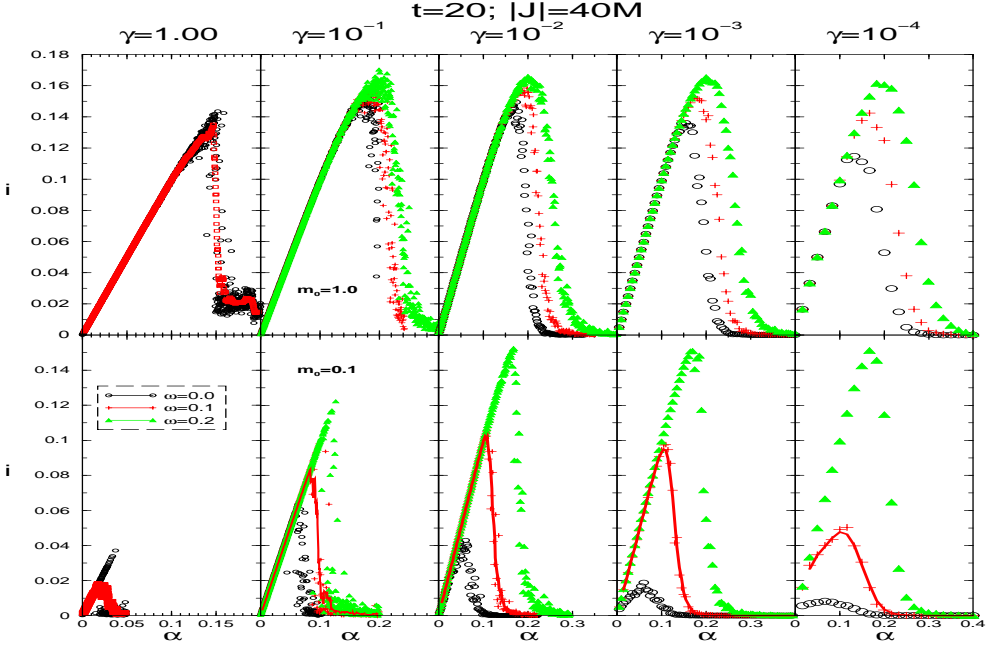}
\caption{\small
The information vs the load,
$i(\alpha)$,
with connectivities from $\gamma=1.0$ (left)
to $\gamma=10^{-4}$ (right).
$N.K=4.10^7$.
In the upper panel,
the simulation starts with $m^0=1.0$,
in the lower panel, with $m^0=0.1$.
Retrieval stops at $t_f=20$.
The randomness are $\omega=0.0$ (open circles),
$\omega=0.1$ (plus) and $\omega=0.2$ (triangles).
The solid line for $\omega=0.1$ with $m^0=0.1$ is a guide to
the eyes.  }
\label{i,agm}
\end{center}
\end{figure}

Each maximal $i_{max}(\gamma;\omega)$ in Fig.\ref{i,agm}
is plotted in Fig.\ref{i,cwm}.
We see that,
for intermediate values of the randomness parameter
$0\leq\omega<0.3$ there is an optimal information respect to
the dilution $\gamma$,
if dynamics is truncated.
We observe that the optimal
$i_{opt}\equiv i_{max}(\gamma_{opt};\omega)$ is shifted to the left
(stronger dilution) when the randomness $\omega$ of the network
increases.

For instance,
with $\omega=0.1$, the optimal is at $\gamma\sim 0.020$
while with $\omega=0.2$, it is $\gamma\sim 0.005$.
This result does not change qualitatively with the
flag time,
but if the dynamics is truncated early,
the optimal $\gamma_{opt}$, for a fixed $\omega$,
is shifted to more connected networks.
However, the behavior depends strongly on the initial condition:
respect to $m_0=0.1$,
where the maximal are pronounced,
with $m_0=1.0$,
the dependence on the topology becomes almost flat.
We see also that for $\omega\geq 0.3$ there is no
intermediate optimal topology.
It is worth to note that the simulation converges to
the theoretical results if $m_0=1.0$ when $t\to\infty$.

\begin{figure}[t]
\begin{center}
\epsfxsize 11.cm 
\epsfysize 8.cm
\epsfbox{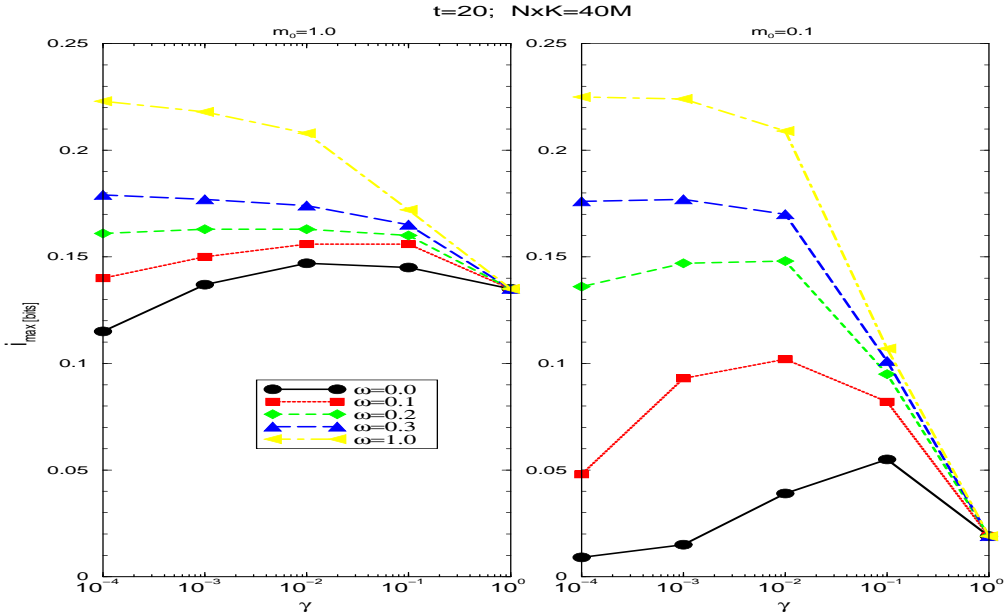}
\caption{\small 
Maximal information $i_{max}=i(\alpha_{max})$ vs $\gamma$,
for simulations with $N.K=4.10^7$,
and several $\omega$.
Initial overlap $m^0=1.0$ (left) and $m^0=0.1$ (right);
the retrieval stops after $t_f=20$ steps. }
\label{i,cwm}
\end{center}
\end{figure}

\subsection{Simulation with Images}

The simulations presented so far use artificial patterns
randomly generated.
In order to check if our results are robust against possibly
correlations existent in realistic patterns,
we test the algorithm with images.
We see that the same non-monotonic behavior for 
$i_{max}(\gamma)$ is observed here.

We have checked the results by using data derived from the
Waterloo image database. 
We are working with square shaped patches.
In order to use Hebb-like non-sparse code binary network 
and still preserve the structure
of the image we process the images preserving the edges,
by applying edge filter.
Each pixel of the patch represents a different neuron. 
The number of connections is up to $N\times K=3\cdot 10^5$ 
and the feasible connectivities
(more than 3 patterns) are $\gamma>0.002$.

Note that the procedure, strictly speaking, 
does not guarantee the conditions for the
distribution of $\xi$, 
because neither $p(\xi=\pm 1)$ is uniform 
(due to the threshold in large blocks),
nor $\xi_i$ are uncorrelated (due to image edges).

We are choosing at random the origin of the patch and the image
to be used from the available 12 images.
The topology of the network is a ring with small world topology.
The results of the simulation,
using Chen filter,
are shown in Fig.\ref{i,gw1}.
The optimal connectivity with $\omega=0.1$ and $t_f=10$ is found to be
$\gamma_{opt}\sim 0.03$.
The fluctuation now are much larger than with random patterns,
due to correlation and small network size.
In the stationary states, $t_f\to\infty$,
the optimal connectivity remains at $\gamma_{opt}\sim 0.03$,
with $i_{opt}\sim 0.165$.
The results agree qualitatively with simulation for random patterns,
Fig.\ref{i,agm},
where the initial overlaps are $m^0=0.1$ and $m^0=1.0$
(in Fig.\ref{i,gw1} it is always $m^0=0.3$).

\section{Conclusions}

In this paper we have studied the dependence of the
information capacity with the topology for an attractor
neural network.
We calculated the mutual information for a Hebb model,
for storing binary patterns,
varying the connectivity ($\gamma$) and randomness ($\omega$) parameters,
and obtained the maximal respect to $\alpha$,
$i_{max}(\gamma,\omega)\equiv i(\alpha_{max};\gamma,\omega)$.
Then we look at the optimal topology, $\gamma_{opt}$
in the sense of the information,
$i_{opt}\equiv i_{max}(\gamma_{opt},\omega)$.
We presented stationary and transient states.
The main result is that larger $\omega$ always leads to
higher information $i_{max}$.

From the stability calculations,
the stationary optimal topology,
is the extremely diluted (RED) network.
Dynamics shows, however, that this is not true:
we found there is an intermediate optimal $\gamma_{opt}$,
for any fixed $0\leq\omega<0.3$.
This can be understood regarding the shape of the attractors.
The ED waits much longer for the retrieval than
more connected networks do,
so the neurons can be trapped in spurious states with
vanishing information.
We found there is an intermediate optimal $\gamma_{opt}$,
whenever the retrieval is truncated,
and it remains up to the stationary states.

Both in nature and in technological approaches to neural devices,
dynamics is an essential issue for information process.
So, an optimized topology holds in any practical purpose,
even if no attemption is payed to wiring or other energetic costs
of random links \cite{AC05}.
The reason is a competition between the broadness
(larger storage capacity)
and roughness (slower retrieval speed)
of the attraction basins.

We believe that the maximization of information respect
to the topology could be a biological criterium
(where non-equilibrium phenomena are relevant)
to build real neural networks.
We expect that the same dependence should happens for
more structured networks and learning rules.

{\bf Acknowledgments}
Work supported by grants TIC01-572, TIN2004-07676-C01-01,
BFI2003-07276,
TIN2004-04363-C03-03
from MCyT, Spain.


\end{document}